\documentclass{article}
\usepackage{emulateapj}
\usepackage{psfig}
\usepackage{apjfonts}
\usepackage{graphicx}
\usepackage{lineno}

\def\jnl@style{\rm}
\def\aaref@jnl#1{{\jnl@style#1}}

\def\aaref@jnl#1{{\jnl@style#1\thinspace}}

\def\aj{\aaref@jnl{{\em AJ}}}                   
\def\araa{\aaref@jnl{{\em ARA\&A}}}             
\def\apj{\aaref@jnl{{\em ApJ}}}                 
\def\apjl{\aaref@jnl{{\em ApJ}}}                
\def\apjs{\aaref@jnl{{\em ApJS}}}               
\def\ao{\aaref@jnl{{\em Appl.~Opt.}}}           
\def\apss{\aaref@jnl{{\em Ap\&SS}}}             
\def\aap{\aaref@jnl{{\em A\&A}}}                
\def\aapr{\aaref@jnl{{\em A\&A~Rev.}}}          
\def\aaps{\aaref@jnl{{\em A\&AS}}}              
\def\azh{\aaref@jnl{{\em AZh}}}                 
\def\baas{\aaref@jnl{{\em BAAS}}}               
\def\jrasc{\aaref@jnl{{\em JRASC}}}             
\def\memras{\aaref@jnl{{\em MmRAS}}}            
\def\mnras{\aaref@jnl{{\em MNRAS}}}             
\def\pra{\aaref@jnl{{\em Phys.~Rev.~A}}}        
\def\prb{\aaref@jnl{{\em Phys.~Rev.~B}}}        
\def\prc{\aaref@jnl{{\em Phys.~Rev.~C}}}        
\def\prd{\aaref@jnl{{\em Phys.~Rev.~D}}}        
\def\pre{\aaref@jnl{{\em Phys.~Rev.~E}}}        
\def\prl{\aaref@jnl{{\em Phys.~Rev.~Lett.}}}    
\def\pasp{\aaref@jnl{{\em PASP}}}               
\def\pasj{\aaref@jnl{{\em PASJ}}}               
\def\qjras{\aaref@jnl{{\em QJRAS}}}             
\def\skytel{\aaref@jnl{{\em S\&T}}}             
\def\solphys{\aaref@jnl{{\em Sol.~Phys.}}}      
\def\sovast{\aaref@jnl{{\em Soviet~Ast.}}}      
\def\ssr{\aaref@jnl{{\em Space~Sci.~Rev.}}}     
\def\zap{\aaref@jnl{{\em ZAp}}}                 
\def\nat{\aaref@jnl{{\em Nature}}}              
\def\iaucirc{\aaref@jnl{{\em IAU~Circ.}}}       
\def\aplett{\aaref@jnl{{\em Astrophys.~Lett.}}} 
\def\apspr{\aaref@jnl{{\em Astrophys.~Space~Phys.~Res.}}}
\def\bain{\aaref@jnl{{\em Bull.~Astron.~Inst.~Netherlands}}} 
\def\fcp{\aaref@jnl{{\em Fund.~Cosmic~Phys.}}}  
\def\gca{\aaref@jnl{{\em Geochim.~Cosmochim.~Acta}}}   
\def\grl{\aaref@jnl{{\em Geophys.~Res.~Lett.}}} 
\def\jcp{\aaref@jnl{{\em J.~Chem.~Phys.}}}      
\def\jgr{\aaref@jnl{{\em J.~Geophys.~Res.}}}    
\def\jqsrt{\aaref@jnl{{\em J.~Quant.~Spec.~Radiat.~Transf.}}}
\def\memsai{\aaref@jnl{{\em Mem.~Soc.~Astron.~Italiana}}}
\def\nphysa{\aaref@jnl{{\em Nucl.~Phys.~A}}}   
\def\physrep{\aaref@jnl{{\em Phys.~Rep.}}}   
\def\physscr{\aaref@jnl{{\em Phys.~Scr}}}   
\def\planss{\aaref@jnl{{\em Planet.~Space~Sci.}}}   
\def\procspie{\aaref@jnl{{\em Proc.~SPIE}}}   

\makeatletter
\DeclareRobustCommand{\Cpp}
{\valign{\vfil\hbox{##}\vfil\cr
   \textsf{C\kern-.1em}\cr
   $\hbox{\fontsize{\sf@size}{0}\textbf{+\kern-0.05em+}}$\cr}%
}
\makeatother

\begin{document}

\bibliographystyle{apj1d}


\title{RADIO AND $\gamma$-RAY CONSTRAINTS ON THE EMISSION GEOMETRY AND BIRTHPLACE OF PSR J2043+2740}
\author{
A.~Noutsos\altaffilmark{1,2,29}, 
A.~A.~Abdo\altaffilmark{3,4}, 
M.~Ackermann\altaffilmark{5}, 
M.~Ajello\altaffilmark{5}, 
J.~Ballet\altaffilmark{6}, 
G.~Barbiellini\altaffilmark{7,8}, 
M.~G.~Baring\altaffilmark{9}, 
D.~Bastieri\altaffilmark{10,11}, 
K.~Bechtol\altaffilmark{5}, 
R.~Bellazzini\altaffilmark{12}
% ...
}
\altaffiltext{1}{Max-Planck-Institut f\"ur Radioastronomie, Auf dem H\"ugel 69, 53121 Bonn, Germany}
\altaffiltext{2}{Corresponding authors: L.~Guillemot, guillemo@mpifr-bonn.mpg.de; A.~Noutsos, anoutsos@mpifr-bonn.mpg.de.}
\altaffiltext{3}{Space Science Division, Naval Research Laboratory, Washington, DC 20375, USA}
\altaffiltext{4}{National Research Council Research Associate, National Academy of Sciences, Washington, DC 20001, USA}
\altaffiltext{5}{W. W. Hansen Experimental Physics Laboratory, Kavli Institute for Particle Astrophysics and Cosmology, Department of Physics and SLAC National Accelerator Laboratory, Stanford University, Stanford, CA 94305, USA}
\altaffiltext{6}{Laboratoire AIM, CEA-IRFU/CNRS/Universit\'e Paris Diderot, Service d'Astrophysique, CEA Saclay, 91191 Gif sur Yvette, France}
\altaffiltext{7}{Istituto Nazionale di Fisica Nucleare, Sezione di Trieste, I-34127 Trieste, Italy}
\altaffiltext{8}{Dipartimento di Fisica, Universit\`a di Trieste, I-34127 Trieste, Italy}
\altaffiltext{9}{Rice University, Department of Physics and Astronomy, MS-108, P. O. Box 1892, Houston, TX 77251, USA}
\altaffiltext{10}{Istituto Nazionale di Fisica Nucleare, Sezione di Padova, I-35131 Padova, Italy}
\altaffiltext{11}{Dipartimento di Fisica ``G. Galilei", Universit\`a di Padova, I-35131 Padova, Italy}
\altaffiltext{12}{Istituto Nazionale di Fisica Nucleare, Sezione di Pisa, I-56127 Pisa, Italy}
% ...
\altaffiltext{29}{Jodrell Bank Centre for Astrophysics, School of Physics and Astronomy, The University of Manchester, M13 9PL, UK}
% ...

\begin{abstract}
We report on the first year of {\em Fermi} $\gamma$-ray observations of pulsed high-energy emission from the old PSR J2043+2740. The study of the $\gamma$-ray efficiency of such old pulsars gives us an insight into the evolution of pulsars' ability to emit in $\gamma$ rays as they age. The $\gamma$-ray lightcurve of this pulsar above 0.1 GeV is clearly defined by two sharp peaks, $0.353\pm 0.035$ periods apart. We have combined the $\gamma$-ray profile characteristics of PSR J2043+2740 with the geometrical properties of the pulsar's radio emission, derived from radio polarization data, and constrained the pulsar-beam geometry in the framework of a Two Pole Caustic and an Outer Gap model. The ranges of magnetic inclination and viewing angle were determined to be $\{\alpha,\zeta\}\sim\{52^\circ$--~$57^\circ,61^\circ$--~$68^\circ\}$ for the Two Pole Caustic model, and $\{\alpha,\zeta\}\sim\{62^\circ$--~$73^\circ,74^\circ$--~$81^\circ\}$ and $\{\alpha,\zeta\}\sim\{72^\circ$--~$83^\circ,60^\circ$--~$75^\circ\}$ for the Outer Gap model. Based on this geometry, we assess possible birth locations for this pulsar and derive a likely proper motion, sufficiently high to be measurable with VLBI. At a characteristic age of 1.2 Myr, PSR J2043+2740 is the third oldest of all discovered, non-recycled, $\gamma$-ray pulsars: it is twice as old as the next oldest, PSR J0357+32, and younger only than the recently discovered PSR J1836+5925 and PSR J2055+25, both of which are at least 5 and 10 times less energetic, respectively.  
\end{abstract}

\section{Introduction}
Before the advent of the {\em Fermi} $\gamma$-ray Space Telescope (hereafter {\em Fermi}), the {\em Compton Gamma Ray Observatory} (CGRO) succeeded in detecting GeV emission from a handful of pulsars, while a much higher number of $\gamma$-ray sources remained unidentified (\nocite{hbb+99}Hartman et al. 1999). It is generally expected that the most energetic pulsars, i.e.~with spin-down luminosity $L_{\rm sd}=4\pi^2I\dot{P}/P^3> 10^{34}$ erg s$^{-1}$ ($I=10^{45}$ g cm$^2$ is the neutron star's moment of inertia; $P$ is the spin period and $\dot{P}$ its first derivative), are the best candidates for detectable $\gamma$-ray emission (\nocite{tbb+99}Thompson et al.~1999). They are typically young pulsars of characteristic spin-down age $\tau_{\rm c}=P/(2\dot{P})<100$ kyr, with $\dot{P}>10^{-15}$ and $P\lesssim 0.1$ s. Those expectations received support with the discovery of the six EGRET pulsars, all of which have $L_{\rm sd}>3\times10^{34}$ erg s$^{-1}$. In that sample, the two oldest pulsars were PSR B1055$-$52, with a spin-down age of $\tau_{\rm c}=535$ kyr, and the radio-quiet Geminga, with $\tau_{\rm c}=342$ kyr; both these pulsars are considered middle-aged amongst the known sample of non-recycled pulsars (e.g.~Fig.~2 in Abdo et al.~2010a\nocite{aaa+10c}).  

The {\em Fermi} satellite was successfully launched on 2008 June 11. During the first six months of the mission, data collected with the Large Area Telescope (LAT) --- the main instrument on-board {\em Fermi} --- were analysed for pulsed $\gamma$ rays from a pre-selected list of energetic radio pulsars (\nocite{sgc+08}Smith et al.~2008). The list of pulsars was selected based on the spin-down luminosity, so that $L_{\rm sd}>10^{34}$ erg s$^{-1}$. Not surprisingly, the vast majority of non-recycled pulsars on the candidate list have $\tau_{\rm c}<10^3$ kyr. One of the few exceptions is the 96-ms pulsar PSR J2043+2740 ($P=0.0961$ s, $\dot{P}=1.23\times10^{-15}$), which is much older than the rest in the sample, with $\tau_{\rm c}=1.2\times 10^3$ kyr. A recent effort to detect a high-energy signal from this pulsar was made with the {\em AGILE} space-telescope by \nocite{ppp+09}Pellizzoni et al.~(2009). They reported a pulsed signature above 50 MeV, at the level of 4.2$\sigma$ above the background. However, a detection was not claimed and the authors calculated a 2$\sigma$ $\gamma$-ray flux upper limit of $F(>100 \ {\rm MeV})<6\times 10^{-8}$ cm$^{-2}$ s$^{-1}$. In addition to the flux estimation, the authors placed an upper limit on the $\gamma$-ray efficiency: $\eta=L_{\gamma}/L_{\rm sd}<0.01$, under the assumption of a 1-sr beam and a spectral index of 2.0. Following the AGILE observations, Abdo et al.~(2010a)\nocite{aaa+10c} folded the first six months of {\em Fermi} data, from the direction of PSR J2043+2740, with the radio-timing ephemeris supplied by Jodrell Bank (\nocite{hlk+04}Hobbs et al. 2004). That analysis yielded the first confident detection (at nearly 5$\sigma$) of $\gamma$-ray pulsations from PSR J2043+2740.    

PSR J2043+2740 was discovered in radio, in the Arecibo millisecond-pulsar survey at 430 MHz (\nocite{trkp94}Thorsett et al. 1994). Based on its dispersion measure, ${\rm DM}=21.0\pm 0.1$ pc cm$^{-3}$ (\nocite{rtj+96}Ray et al.~1996), and the NE2001 free-electron density model of Cordes \& Lazio (2002)\nocite{cl02}, the distance estimate for this pulsar is $D\approx 1.8$ kpc (\nocite{mhth05}Manchester et al.~2005). PSR J2043+2740 lies near the south-western shell of the Cygnus Loop ($\sim 15$ pc outside the observable edge), perhaps suggesting an association with the remnant. However, the evidence so far suggests that such an association is unlikely: the distance to the Cygnus Loop has been estimated to $540^{\, +100}_{\, -80}$ pc (Blair et al.~2005\nocite{bsr05}; Blair et al.~2009\nocite{bstc+09}); in addition, assuming that the pulsar was born within the observable limits of the remnant, the age of the latter ($<12$ kyr; \nocite{sb02}Sankrit \& Blair~2002) suggests a transverse velocity of $>980$ km s$^{-1}$ for the pulsar, which is significantly higher than the average birth velocity of the known pulsar sample ($400\pm 40$ km s$^{-1}$; \nocite{hllk05}Hobbs et al.~2005). Last but not least, the pulsar's characteristic age, as calculated from its spin parameters, is two orders of magnitude higher than the remnant's. Therefore, these discrepancies need to be reconciled before an association can be claimed. 

There are indeed examples of age discrepancy between pulsars and their associated remnants, like in the case of PSR J0538+2817 and the supernova remnant S147 (\nocite{acj+96}Anderson et al.~1996). This pulsar has a characteristic spin-down age of 620 kyr but has been confidently associated with the 40-kyr remnant via pulsar timing and Very Long Baseline Interferometry (VLBI; \nocite{rn03}Romani \& Ng 2003; \nocite{klh+03}Kramer et al.~2003; \nocite{nrb+07}Ng et al.~2007). Those VLBI measurements, combined with previous X-ray observations of the pulsar's thermal profile (\nocite{mkz+03}McGowan et al.~2003) have revealed a hot neutron star with a transverse velocity of 400 km s$^{-1}$, which matches the observed average. The above studies concluded that PSR J0538+2817 must have been born with a slow initial spin period, very close to that observed today, thus invalidating the usual assumption of a short birth period for this pulsar.   

The properties of PSR J2043+2740 make it an intriguing pulsar for high-energy studies with {\em Fermi}: it is one of the shortest-period, non-recycled $\gamma$-ray pulsars without a known SNR association. Indeed, there are a number of recent {\em Fermi} detections of non-recycled pulsars that have shorter periods than PSR J2043+2740 and for which there is yet no association with a remnant: i.e.~PSRs J1028$-$5819, J1718$-$3825, J1420$-$6048, J1813$-$1246 (Abdo et al.~2010a\nocite{aaa+10c}). However, these pulsars are $\sim 10^2$ times younger than PSR J2043+2740 and are much more likely to be associated with a nearby remnant; on the contrary, if the characteristic age for PSR J2043+2740 corresponds to its true age (see section~\ref{subsec:grayeff}), then it is very unlikely that there would be any visible remnant left for an association to be possible. Moreover, the fast rotation of PSR J2043+2740 implies a relatively high spin-down luminosity ($L_{\rm sd}=5.6\times 10^{34}$ erg s$^{-1}$) compared to pulsars of similar characteristic age. This means that, in terms of energetics, PSR J2043+2740 is on a par with much younger pulsars, like PSR J1835$-$0643, which is an order of magnitude younger. 

Apart from arousing observational interest, this pulsar's physical properties are also attractive in theoretical investigations. In terms of $\gamma$-ray observability, $L_\gamma/D^2$, PSR J2043+2740 has been previously considered as a strong candidate for $\gamma$-ray emission, in the framework of both Polar Cap (PC) and Outer Gap (OG) models (\nocite{rd98}Rudak \& Dyks 1998; \nocite{hib02}Hibschman 2002; \nocite{cz98}Cheng \& Zhang 1998; \nocite{mc00}McLaughlin \& Cordes 2000). It should be noted, however, that the majority of {\em Fermi} pulsar observations to date have produced $\gamma$-ray spectra that disagree with the predicted super-exponential cutoffs of PC models (e.g.~\nocite{aaa+09c}Abdo et al.~2009; \nocite{aaa+10a}Abdo et al.~2010b). Nevertheless, even within the framework of outer-magnetospheric models, there exist alternative emission geometries to the traditional OG that describe the production of $\gamma$ rays at high altitudes above the pulsar surface: one such geometry is that of the Two-Pole Caustic Model (TPC; \nocite{dr03}Dyks \& Rudak 2003). The detection of PSR J2043+2740 makes it possible to test the predictions of the above models for this pulsar against the properties derived from observations. In section \ref{subsec:HEmodels}, we discuss the results from the derived emission geometry from radio polarization data and the $\gamma$-ray lightcurves, and what those suggest for the model describing this pulsar's emission.   

Furthermore, the measurement of the $\gamma$-ray efficiency, $\eta=L_{\gamma}/L_{\rm sd}$, for the old PSR J2043+2740 extends the studies of $\eta$ by a factor of 2 in characteristic age. This can help us confirm or reject previous claims for an increasing $\gamma$-ray efficiency with pulsar age (\nocite{buc80}Buccheri 1980; \nocite{har81a}Harding 1981; Zhang \& Cheng 1998\nocite{zc98}). A discussion on this subject can be found in section \ref{subsec:grayeff}.  

The present article reports on the results of our analysis of 14 months of {\em Fermi} data from the direction of PSR J2043+2740. The LAT instrument on-board {\em Fermi} is sensitive to $\gamma$ rays of energies from 0.02 to 300 GeV; its sensitivity is an order of magnitude higher than EGRET and more than 3 times higher compared to {\em AGILE} above 0.5 GeV (\nocite{aaa+09s}Atwood et al.~2009). Using the data collected with LAT during the first 14 months of operation (2008 Aug 4 -- 2009 Oct 17), we have detected $\gamma$-ray pulsations at a very high significance ($\approx 7\sigma$) from PSR J2043+2740. This work strengthens the previously published detection of this pulsar with {\em Fermi} data (\nocite{aaa+10c}Abdo et al.~2010a). 

\section{Radio Observations and Data Analysis}
The analysis of radio data from PSR J2043+2740 consisted of two parts: (a) the analysis of timing data from Jodrell Bank observations at 1.4 GHz, from which we derived a rotational ephemeris that was used to fold the $\gamma$-ray data; and (b) the analysis of polarimetric data from observations at 1.4 GHz with the Effelsberg radio telescope, which we used to determine the beam geometry of the pulsar's emission.

\subsection{Timing Data and Analysis}
The rotational ephemeris for PSR J2043+2740, which is required to fold the $\gamma$-ray photons with the pulsar's period, was obtained from regular timing observations with the Lovell 76-m radio telescope, at 1.4 GHz. The Jodrell Bank radio observatory has been timing this pulsar regularly since 1996, as part of its extensive pulsar-timing program (\nocite{hlk+04}Hobbs et al. 2004). In our analysis, we used timing data that spanned from 2008 June 17 to 2009 October 17, covering 78, 12-minute integrations of the radio pulses, from each of which a precise time of arrival (TOA) was determined. The set of TOAs was analysed with the pulsar-timing package TEMPO2 (\nocite{hem06}Hobbs et al. 2006), which we used in order to fit a timing model to the TOAs, describing the pulsar rotation during the above range of dates. The only free parameters of the timing model were the pulsar spin-frequency and its first two time derivatives, while the pulsar position and DM were deemed sufficiently well-determined from previous observations. After fitting, the pulsar ephemeris was a good description of the pulsar rotation, with the RMS of the timing residuals being $\approx$ 120 $\mu$s across the entire fitted range. 
The uncertainty in DM results in a possible error in the time delay between the radio and $\gamma$-ray frequencies (with $f_{\rm LAT}\rightarrow\infty$), via the dispersion law, $\delta t=(4.15\times10^6 \ {\rm ms})\times \sigma_{DM}/f^2$. Substituting for the DM uncertainty and the radio frequency gives $\delta t\approx 0.21$ ms, which corresponds to only 0.2\% phase error; this is of the same order of magnitude as the uncertainty of the ephemeris. However, given the low number of $\gamma$-ray events collected from PSR J2043+2740, so far, the impact of both errors on the pulsar's $\gamma$-ray lightcurve is negligible.  

The timing parameters used in this study will be made available on the servers of the {\em Fermi} Science Support Center (FSSC)\footnote{http://Fermi.gsfc.nasa.gov/ssc}

\subsection{Polarimetric Data and Analysis}
\label{subsec:poldata}
In addition to the timing observations, we performed polarization observations of PSR J2043+2740 at 1.4 GHz, with the Effelsberg 100-m radio telescope. The 21cm receiver of the Effelsberg observatory is equipped with left and right circular polarization feeds, with a system equivalent flux density of $\approx 20$ Jy. The two polarization channels were fed into a Digital Filterbank back-end, which was set to de-disperse the 150 MHz of available bandwidth, split into 1024 frequency channels. For calibration purposes, during the entire 45-minute observation we triggered the receiver's noise diode at the pulsar frequency but separated by 0.5 periods from the pulse position. The noise diode injects an artificial, pulsed signal that is 100\% linearly polarized, and which can be used as a reference source for correcting for the gain differences between the polarization feeds, in the post-processing. 

\begin{figure*} 
\vspace*{10pt} 
\includegraphics[scale=0.45]{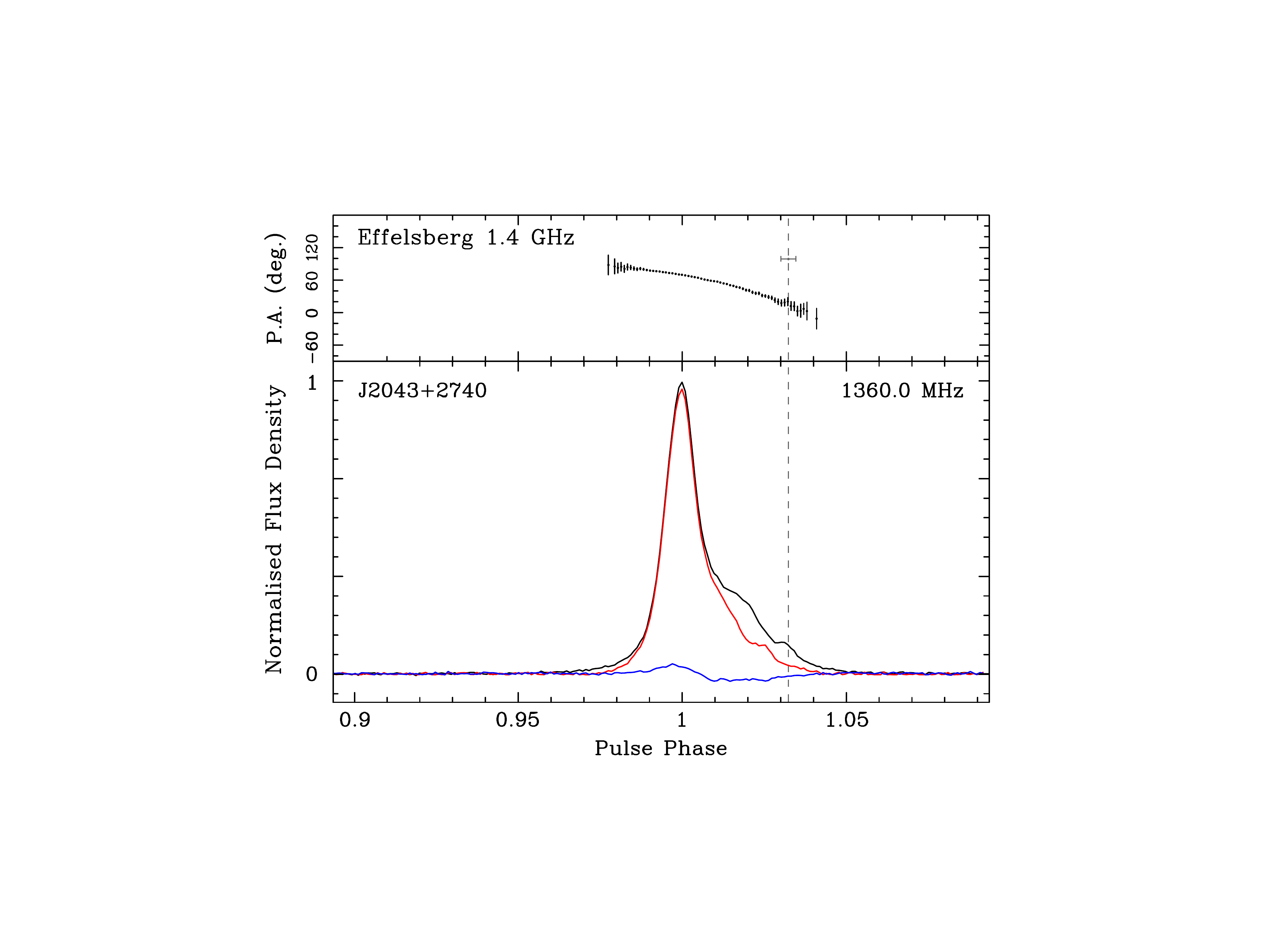}
\caption{\label{fig:polprofile} Polarization profile of PSR J2043+2740, at 1.4 GHz, from a 45-minute integration with the Effelsberg radio telescope. Bottom panel: profiles of the total intensity (black, solid line), linearly polarized intensity (red line) and circularly polarized intensity (blue line). Top panel: PA profile (black error bars). The grey dashed line denotes the rotational phase at the minimum approach of the line-of-sight to the magnetic pole (i.e.~$\phi_0$). The uncertainty on $\phi_0$ is shown with a horizontal error-bar. }
\end{figure*}

We used the PSRCHIVE software package (\nocite{hvm04}Hotan et al.~2004) to calibrate the Effelsberg data and produce a high-signal-to-noise (${\rm s/n}$) polarization profile of PSR J2043+2740. The total-intensity, linear- and circular-polarization profiles of this pulsar are shown in Fig.~\ref{fig:polprofile}. Furthermore, in the top panel of the same figure we show the polarization position angle (hereafter PA) profile across the pulse, given by ${\rm PA}=0.5\arctan(U/Q)$, where $Q$ and $U$ are the Stokes parameters of the linearly polarized emission at each pulse longitude: only PAs with ${\rm s/n}>4$ are shown in the plot. 

Adopting a standard Rotating Vector Model (RVM), described in Radhakrishnan \& Cooke (1969), we determined the model-parameter space from fits to the observed PAs across the pulse: i.e.~$\alpha$, the inclination angle between the magnetic and rotating axes; $\zeta=\alpha+\beta$, the viewing angle, where $\beta$ is the minimum angular separation between the observer's line-of-sight and the magnetic axis, which occurs at rotational phase $\phi_0$; and ${\rm PA}_0$, the PA at $\phi_0$. As is the case with many other polarized pulsars, the PA profile of PSR J2043+2740 reveals to us only part of the full swing of the magnetic axis across our field of view. Therefore, there are large uncertainties on the fitted parameters. In order to estimate the confidence intervals of $\alpha$ and $\zeta$, we fitted the PAs with an RVM model: we used a $500\times 500$ grid in $\alpha=\{0^\circ$--~$180^\circ\}$ and $\zeta=\{0^\circ$--~$180^\circ\}$ and left $\phi_0$ and $\psi_0$ as free parameters to be fitted with a $\chi^2$-minimization procedure. Each grid point was then assigned to the calculated $\chi^2$ value from the respective fit and 1, 3 and $5\sigma$ contours were generated.

Despite the aforementioned polarization calibration procedure, which largely corrects for the parallactic angle changes during the observation and the frequency-dependent gain differences between the dipoles (for a description see Johnston 2002\nocite{joh02}), the PAs that we measured with our back-end are subjected to additional instrumental rotations by the different components between the receiver and the back-end. The instrumental rotation is assumed to be time and frequency independent, and one can account and correct for this constant shift in the PAs by observing with the same setup a set of highly polarized pulsars with well-known polarization properties. Given that these pulsars are also calibrated, the difference between the measured PAs (corrected to infinite frequency) and previously published absolute PA values should correspond to the instrumental rotation. Application of the measured instrumental rotation to the calibrated profile of any pulsar observed with the same system should result in absolute PAs.

We performed polarization observations of 5 pulsars --- PSR~B0355+54, PSR~B0740$-$28, PSR~B1929+10, PSR~B1953+50 and PSR~B2154+40 --- with the same receiver setup and produced calibrated PA profiles across the pulse. Using the calibrated PAs, we measured the rotation measures (RM) of these pulsars by fitting for the PA rotation across our 150 MHz band. Following that, we used the resulting RMs to correct the measured PAs to infinite frequency. Then, we calculated the mean difference between the published values of ${\rm PA}_0$ and those from our measurements in our sample of 5 pulsars. Absolute polarization profiles for these pulsars were found in Johnston et al.~2005\nocite{jhv+05} and Carr 2007\nocite{car07}. The calculated $\Delta{\rm PA}$ between the measured and published PAs was ${\rm PA}_{\rm meas}-{\rm PA}_{\rm pub}={\rm 13\fdg6}\pm {\rm 2\fdg8}$.

\begin{figure*}
\begin{center}
\includegraphics[scale=0.45]{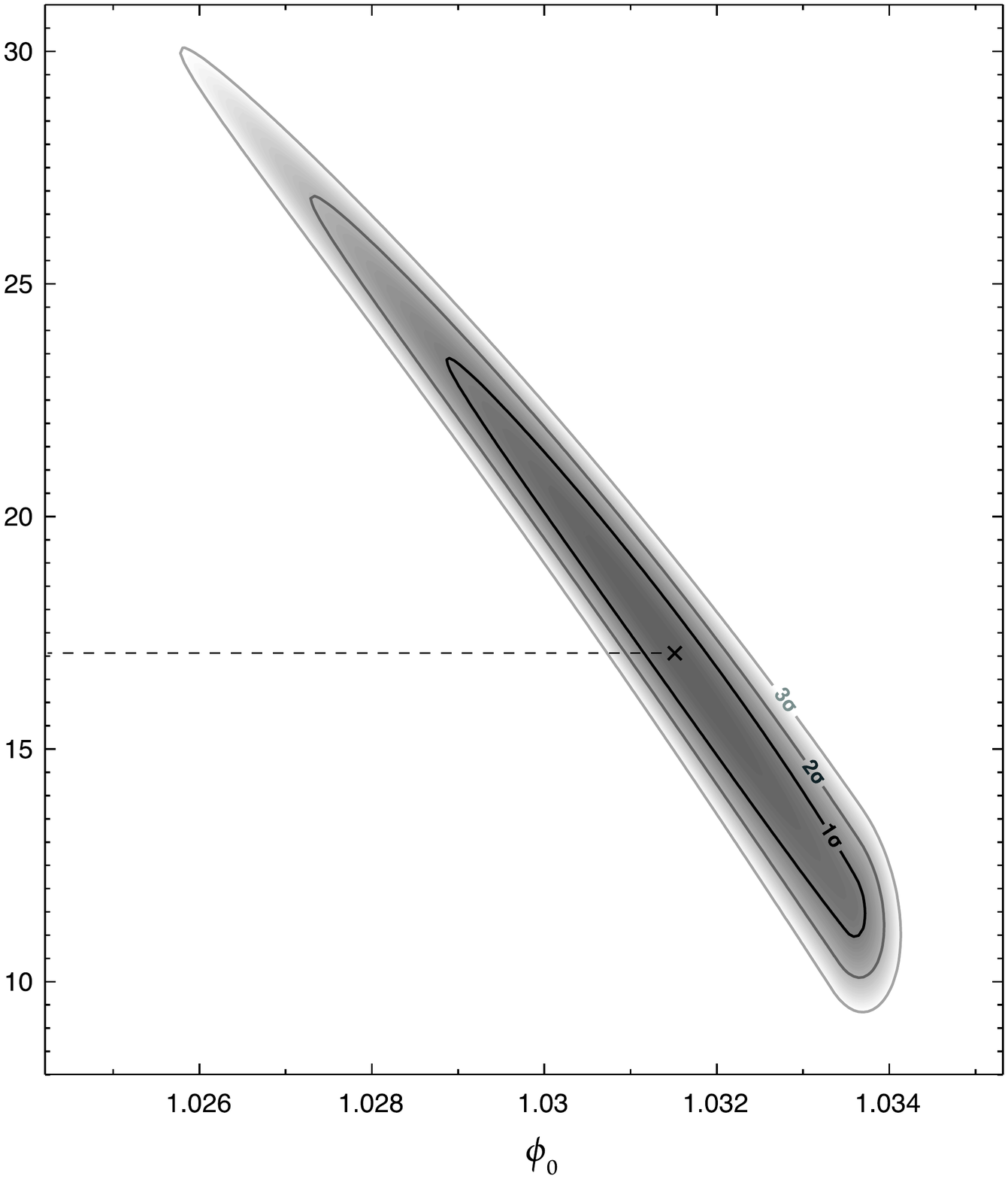}
\caption{\label{fig:confcont} Confidence contours for $\phi_0$ and ${\rm PA}_0$ from fits to the polarization data of Fig.~\ref{fig:polprofile} with an RVM model, where $\alpha$ and $\zeta$ in the model were taken from a $500\times500$ grid covering 0$^\circ$ -- $180^\circ$ in each parameter. The best-fit pair of $\phi_0$ and ${\rm PA}_0$, corresponding to $\chi^2_{\rm r}\approx 0.8$, is shown with a cross. }
\end{center}
\end{figure*}

For PSR~J2043+2740, following the above procedure, we measured ${\rm RM}=-92.7\pm 1.9$ rad m$^{-2}$ and corrected the PAs to infinite frequency, according to that value. Finally, by subtracting $\Delta{\rm PA}$ from all measured PA values, we calculated the absolute polarization-angle profile shown in the top panel of Fig.~\ref{fig:polprofile}. The RVM-model contours from the radio-polarization data allowed us to place constraints on the ranges of the $\alpha$ and $\zeta$ parameters (see section \ref{subsec:HEmodels}). Following the aforementioned RVM-fitting approach but, instead, using a grid in $\phi_0$ and ${\rm PA}_0$ and fitting for $\alpha$ and $\zeta$, we were able to generate confidence contours for $\phi_0$ and ${\rm PA}_0$ (Fig.~\ref{fig:confcont}). The best solution ($\chi^2_{\rm r}\approx 0.8$) gave $\phi_0=1.032\pm 0.002$ and ${\rm PA}_0=17^\circ\pm6^\circ$.

\section{$\gamma$-Ray Observations and Data Analysis}

\subsection{Lightcurve analysis}
\label{subsec:lighcurv}
The raw event data from the {\em Fermi} LAT observations were subjected to cuts, in order to reduce the $\gamma$-ray background. We kept the ``diffuse'' class events as defined under the P6\_V3 instrument response function (IRF), having the best probability of being $\gamma$-ray photons (Atwood et al 2009), and rejected all events with zenith angles greater than 105$^\circ$ in order to reduce the contribution of the Earth's $\gamma$-ray albedo\footnote{Earth's albedo of $\gamma$ rays is produced by cosmic rays interacting with the atmosphere.}. We kept events within an energy-dependent radius corresponding to 68\% of the \emph{Point Spread Function} (PSF) of the LAT, from the position of J2043+2740 ($\alpha_\mathrm{J2000} = {\rm 310\fdg931}$, $\delta_\mathrm{J2000} = {\rm 27\fdg682}$), $\theta_{68} = {\rm 0\fdg8} \times E^{-0.8}$ with $E$ in GeV, with a minimum value of 0\fdg1 for $E > 10$ GeV. A maximum radius of $1^\circ$ around the pulsar was imposed to remove as many low-energy $\gamma$ rays due to the Galactic diffuse emission as possible. The remaining 1244 photons were phase-folded using the radio ephemerides described above, and the \emph{Fermi} plug-in provided by the LAT team and distributed in the TEMPO2 pulsar timing package (\nocite{hem06}Hobbs et al. 2006).

\begin{figure}[htbp]
\begin{center}
\includegraphics[scale=0.67]{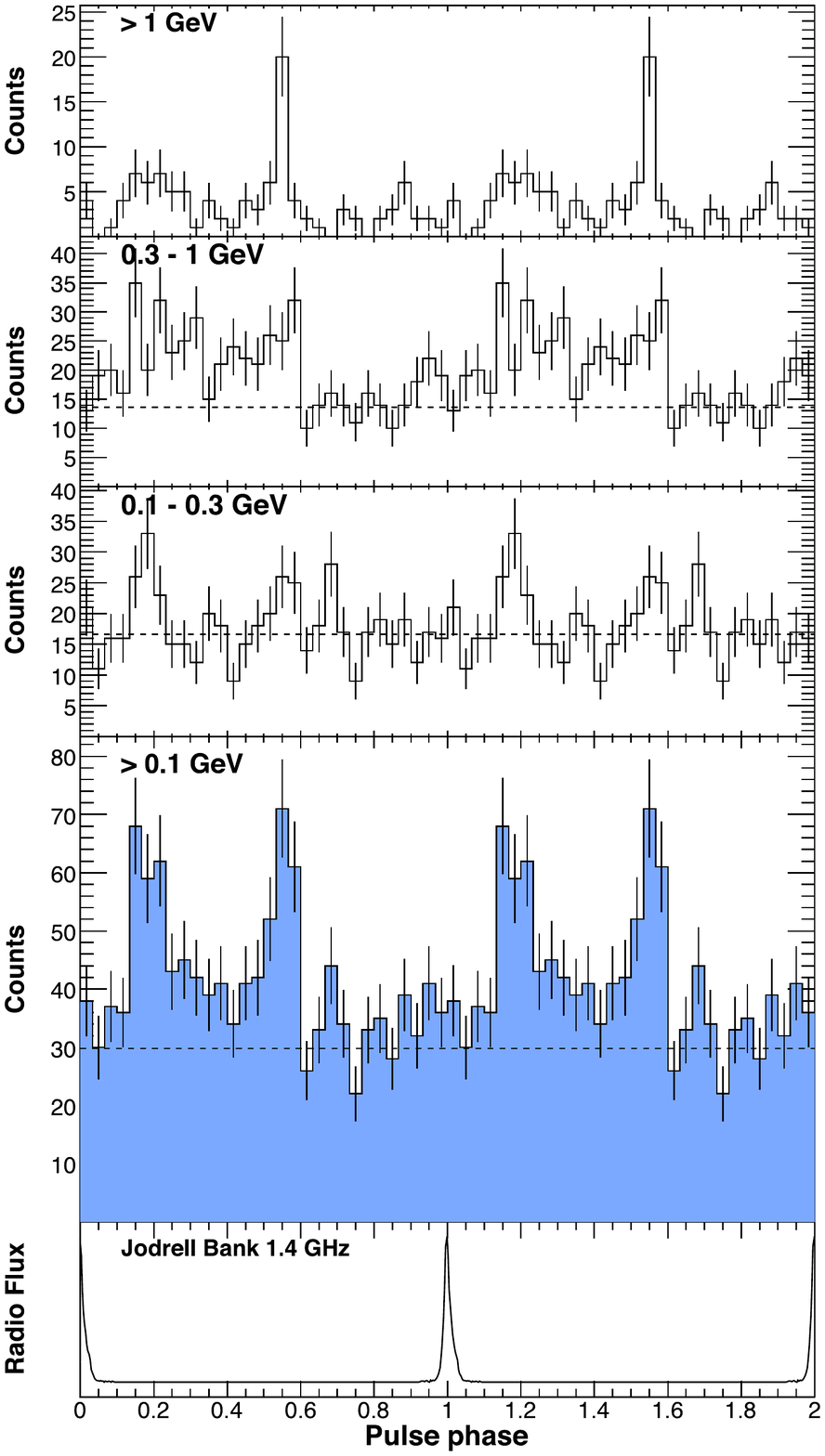}
\caption{\label{fig:lightcurves} Top four panels: $\gamma$-ray lightcurves of PSR J2043+2740 for $E>0.1$ GeV, $0.1<E<0.3$ GeV, $0.3<E<1$ GeV and $E>1$ GeV. Bottom panel: radio profile of PSR J2043+2740 from Jodrell Bank observations at 1.4 GHz. The horizontal dashed lines show the background levels estimated from a surrounding annulus.}
\end{center}
\end{figure}

Fig.~\ref{fig:lightcurves} shows the generated lightcurves for all events above 0.1 GeV, those between 0.1 and 0.3 GeV, 0.3 and 1 GeV, and all events above 1 GeV. The highest-energy event in the data after the application of cuts was 4.9 GeV, and falls at 0.30 in phase. The bottom plot, in the same figure, shows the radio profile of PSR J2043+2740 from Jodrell Bank observations at 1.4 GHz. All profiles were phase-aligned to the maximum of the radio emission (set by definition to $\phi=0$), to allow a comparison study between the longitude of radio and high-energy emission. The overall lightcurve above 0.1 GeV shows two significant peaks of $\gamma$-ray emission, which is in agreement with previous \emph{Fermi} observations (\nocite{aaa+10c}Abdo et al.~2010a), but departs from the marginally significant lightcurve recorded by AGILE (\nocite{ppp+09}Pellizzoni et al.~2009). Application of the binning-independent $H$-test for uniformity (\nocite{dsr89}de Jager et al.~1989) to those events yielded $H = 69.5$, which translates to a $\approx 7\sigma$ detection (de Jager \& B\"usching 2010\nocite{db10}). Inspection of the remaining histograms showed that the phase region defining P1 in the aggregate lightcurve is mainly populated by $\gamma$ rays coming from the lower-energy windows, i.e.~0.1--1 GeV. In contrast, the $\gamma$ rays that form P2 in the total lightcurve mainly come from the highest-energy window, i.e.~$>1$ GeV. We therefore conclude that there is a spectral dependence of the $\gamma$-ray lightcurve, with the second peak harder than the first one.


In order to accurately determine the peak positions in the lightcurve of PSR J2043+2740, a $C$-test (\nocite{dej94}de Jager 1994) was applied. This test, unlike the $H$-test, is sensitive to phase and pulse-width information, which allows one to maximize the significance of the signal by scanning in phase and pulse width around the roughly known pulse positions. A grid of $200\times 60$ trials of peak position, $\phi$, and pulse width, $W$, respectively, was generated and a probability of random occurrence was calculated for each point on the grid. The pulse width was defined as the FWHM (Full Width at Half Maximum) of a Gaussian pulse shape. 
From the test, we found that $\phi_{\rm P1} = 0.20 \pm 0.03$ and $W_{\rm P1}=0.19 \pm 0.06$, and $\phi_{\rm P2} = 0.55\pm 0.02$ and $W_{\rm P2}=0.06\pm0.01$ describe well the peak position and width.

From the above values we derive a phase lag between the maximum of the radio emission peak and the first $\gamma$-ray peak of $\delta = 0.20 \pm 0.03$, and a separation between the two $\gamma$-ray peaks of $\Delta = 0.35 \pm 0.04$, in agreement with previously released results (Abdo et al.~2010a\nocite{aaa+10c}). 

\subsection{Spectral analysis}
\label{subsec:spectral}


The phase-averaged spectrum for PSR J2043+2740 was obtained by performing a maximum likelihood analysis (\nocite{mbc+96}Mattox et al. 1996) of the LAT data within 15$^\circ$ from the pulsar, using the standard tool ``gtlike''. The Galactic diffuse emmission was modeled using the gll\_iem\_ v02 map cube. The extragalactic diffuse and residual instrument background components were modeled jointly using the isotropic\_iem\_v02 template. Both models are available for download with the {\em Fermi} Science Tools\footnote{http://fermi.gsfc.nasa.gov/ssc/data/analysis/scitools/overview.html}. All sources within 15$^\circ$ from J2043+2740 in the 11-month catalog of point sources were included in the analysis, and parameters for sources more than 5$^\circ$ away from the pulsar were fixed in the fit (\nocite{aaa+10d}Abdo et al.~2010c). We modeled the emission from J2043+2740 as a simple power law and an exponentially cut-off power law of the following form:

\begin{eqnarray}
\frac{dN}{dE} = N_0 \left( \frac{E}{\mathrm{1 \ GeV}} \right)^{-\Gamma} {\rm e}^{-\frac{E}{E_{\rm c}}}
\label{expcutoff}
\end{eqnarray}

In this expression, $\Gamma$ is the power-law index, $E_{\rm c}$ is the cut-off energy, and $N_0$ is a prefactor term. The exponentially cut-off power-law emission model is preferred at the $5.5\sigma$ level over the simple power-law model. We obtain $N_0 = (1.2 \pm 0.2) \times 10^{-8}$ photons cm$^{-2}$ s$^{-1}$ GeV$^{-1}$, $\Gamma = 1.1 \pm 0.2$ and $E_{\rm c} = 0.8 \pm 0.1$ GeV. We also modeled the pulsar with a super-exponentially cut-off power law of the form $N_0(E/1 \ {\rm GeV})^{-\Gamma}\exp[-(E/E_{\rm c})^\beta]$, where $\beta$ was left free in the fit. We measured $\beta = 1.0 \pm 0.4$, and therefore we conclude that the simple exponential cut-off power law models the present data well. 

Integrating Equation (\ref{expcutoff}) for energies above 0.1 GeV gives an integral photon flux $F_{\mathrm{> 0.1\ GeV}} = (2.2 \pm 0.4) \times 10^{-8}$ photons cm$^{-2}$ s$^{-1}$, and an energy flux $G_{\mathrm{> 0.1\ GeV}} = (1.5 \pm 0.2) \times 10^{-11}$ erg cm$^{-2}$ s$^{-1}$. The quoted errors are statistical only. The effect of the systematic uncertainties in the effective area on the spectral parameters is $\delta\Gamma=(+0.3, -0.1)$,	$\delta E_{\rm c}=(+20\%, -10\%)$, $\delta F_{\mathrm{> 0.1\ GeV}}=(+30\%, -10\%)$, and $\delta G_{\mathrm{> 0.1\ GeV}}=(+20\%, -10\%)$ (Abdo et al.~2010a\nocite{aaa+10c}). These systematic uncertainties were assessed by making the analysis as described above, but using an effective area for the LAT modified by $\pm$ 10\% at 0.1 GeV, $\pm$ 5\% at 0.5 GeV, and $\pm$ 20\% at 10 GeV with linear extrapolations in log space between. Our values for $\Gamma$, $E_{\rm c}$, $F_{\mathrm{> 0.1\ GeV}}$ and $G_{\mathrm{> 0.1\ GeV}}$ are in good agreement with those of Abdo et al.~(2010a)\nocite{aaa+10c}, but with improved statistical uncertainties.

\section{Discussion}

\subsection{High-Energy Models}
\label{subsec:HEmodels}
All recent detections by {\em Fermi} suggest that the observed $\gamma$-ray emission from pulsars is distributed over a large fraction of the celestial sphere: i.e.~$f_\Omega\gtrsim 1$, where $f_\Omega$ is a flux-correction factor for the fact that the actual phase-averaged flux of a pulsar, integrated over the whole celestial sphere, may be higher ($f_\Omega>1$) or lower ($f_\Omega<1$) than would be inferred from assuming isotropic emission based on the phase-averaged flux for the Earth line-of-sight (\nocite{wrwj09}Watters et al.~2009). 

The phase-averaged flux integrated over the whole sky is a measure of the size of the emission region in the pulsar magnetosphere. In general, values of $f_\Omega$ greater than 1 are consistent with the ``fan-beam'' emission from outer-magnetospheric gaps, whereas PC models tend to produce narrow beams, i.e.~$f_\Omega\ll 1$.

\begin{figure*} 
\begin{center}
\vspace*{10pt} 
\includegraphics[scale=0.45]{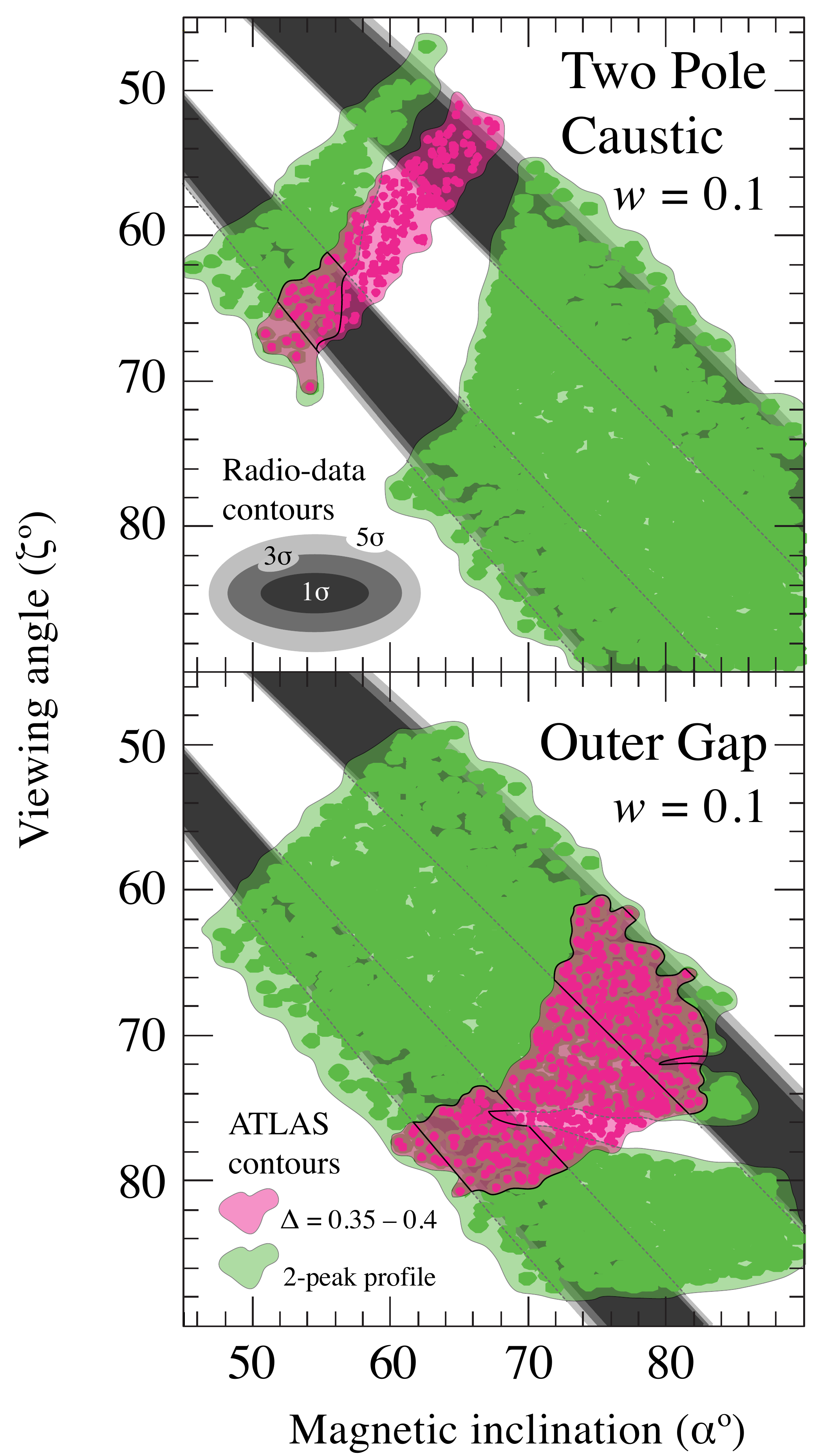}
\caption{\label{fig:atlas} Plots of the magnetic inclination, $\alpha$, versus the viewing angle, $\zeta$, for $\gamma$-ray pulsars with Two Pole Caustic emission regions (top) and Outer Gap regions (bottom), and with an assumed gap thickness of the emission region of $w\sim 0.1$. In these plots, the allowed geometries from the ATLAS maps of Watters et al.~(2009) are shown with green and pink contours: the green contours correspond to a $\gamma$-ray profile with two major peaks; the pink contours, to a profile with phase separation of 0.35--0.4 pulse periods between the major peaks. In both models, the geometry is further constrained by the greyscale contours that were derived from RVM fits to the radio polarization data of PSR J2043+2740. The overlapping regions between the ATLAS and radio contours are delineated with solid, black lines.}
\end{center}
\end{figure*}

The geometrical constraints from our radio polarization measurements can be combined with information derived from the observed $\gamma$-ray lightcurve, to place more stringent limits on the allowed values of $\alpha$ and $\zeta$. More specifically, we can use the measured P1--P2 peak separation of $\Delta \approx 0.35$ and the gap-thickness value for PSR J2043+2740, $w=(L_{\rm sd}/10^{33} \ {\rm erg \ s}^{-1})^{-1/2}\sim 0.1$, defined in Watters et al.~(2009), to explore the allowed geometries that are consistent with the TPC and the OG model, based on the ATLAS maps of Watters et al.~(2009). Fig.~\ref{fig:atlas} shows the $\alpha$ and $\zeta$ values allowed by the RVM fit of the PA in greyscale contours. The green and pink points show the allowed geometries from the ATLAS of Watters et al.~(2009), for a two-peaked gamma-ray profile and for a profile with a phase separation of 0.35--0.4 between the major peaks, respectively. The regions where all contours (from the radio data and from the $\gamma$-ray models) overlap are delineated with solid, black lines. 

For the TPC model, the overlap between the contours covers roughly the ranges
$\alpha \sim$ $52^\circ$--~$57^\circ$ and $\zeta \sim 61^\circ$--~$68^\circ$. For the OG model, there are two separate regions of overlap between the radio contours and the ATLAS maps: these are roughly $\alpha \sim$ $62^\circ$--~$73^\circ$ and $\zeta \sim 74^\circ$--~$81^\circ$; and $\alpha \sim$ $72^\circ$--~$83^\circ$ and $\zeta \sim 60^\circ$--~$75^\circ$. The presence of overlapping regions in both TPC and OG contour plots does not allow us to exclude either of the two models. However, it is worth noting that the two outer peaks expected in the TPC geometry both have progressive leading edges and sharp trailing edges, whereas the OG model predicts symmetric peaks (see Appendix of \nocite{wrwj09}Watters et al.~2009). Therefore, the gamma-ray lightcurve shape we observe is consistent with the OG model prediction and inconsistent with the TPC.

Furthermore, as was mentioned earlier, the P1/P2 ratio decreases with energy. There are already several investigations of the energy dependence of P1 and P2 for Fermi pulsars (\nocite{aaa+10e}Abdo et al.~2010e; \nocite{aaa+10f}Abdo et al.~2010f; \nocite{aaa+10g}Abdo et al.~2010g). In all those cases, a double-peaked profile is observed that reveals a harder spectral index for P2 compared to that of P1. This indicates, for curvature radiation-dominated models, a higher accelerating electric field and/or a smaller radius of field-line curvature in the P2 emission region; future modeling can use these spectral variations to probe the underlying physics. Interestingly, the highest-energy $\gamma$ ray is found nearer to P1 than to P2, which means that it is either part of the surviving photons from the respective cut-off process operating near the P1 region, or simply a background photon not associated with the pulsar.

\subsubsection{Altitude of Emission}

An interesting feature that emerges from the radio-polarization profile of PSR J2043+2740 is the shape of the circularly polarized flux, which seems to change handedness very near the peak of the total flux (it lags the latter by $\approx$ 0.6\% of the pulse period). Such features in pulsar profiles have been associated with emission from near the magnetic pole, viewing angles almost along the local magnetic field (Rankin 1986\nocite{ran86}). Interestingly, the RVM fit for this pulsar gives a $\phi_0$ that lags the V-swing by $\approx 2$\% of the pulse period. According to the original relativistic model of Blaskiewicz, Cordes \& Wasserman (1991)\nocite{bcw91}, for which Dyks (2008)\nocite{dyk08} provided recently a simple explanation in terms of the relative acceleration of the corotating pulsar magnetosphere to the observer's reference frame, such a difference between the fiducial phase of an RVM swing and that of the profile is expected if the emission is altitude-dependent. In that case, the radio emission would be generated from an altitude that is roughly 2\% the size of the light-cylinder ($R_{\rm LC}\sim 4.59\times 10^4$ km).
  
On the other hand, one can place limits on the altitude of the $\gamma$-ray emission based on the highest observed photon energy from PSR J2043+2740. The most energetic $\gamma$ ray in the pulsar's lightcurve corresponds to an energy of $\epsilon_{\rm max}=4.9$ GeV. In the framework of a standard PC model, \nocite{bar04}Baring (2004) placed a lower limit on the radius of the high-energy emission, based on $\epsilon_{\rm max}$. The minimum radius (from the neutron star's center), based on the $\gamma$-$B$ absorption of $\gamma$ rays propagating through the magnetosphere, was estimated to be $r\gtrsim (\epsilon_{\rm max}B_{12}/1.76 \ {\rm GeV})^{2/7}P^{\: -1/7}R_{\star}$, where $R_{\star}$ is the neutron star radius. Substituting for $P=0.0961$ s and $B_{12}=B/(10^{12} \ {\rm G})=0.354$, the above inequality yields $r\gtrsim 1.39 R_{\star}$, which rules out emission models that produce $\gamma$ rays very near the polar caps. Recently, Lee et al.~(2010)\nocite{ldw+10} performed a largely similar alternative analysis of the altitude of gamma-ray emission, employing a well-known high-energy asymptotic form for the magnetic pair creation rate. They concluded that the minimum emission height can be approximated with $r\gtrsim 0.11\times (\epsilon_{\rm max}B_{12})^{2/5}P^{\: -1/5}R_{\star}$. Substituting the values of PSR J2043+2740, we get $r\gtrsim 3.5R_{\star}$, which again shows that the highest-energy emission from this pulsar is not produced near the polar caps.

\subsection{$\gamma$-Ray Efficiency}
\label{subsec:grayeff}
As was mentioned in the introduction, the high characteristic age of PSR J2043+2740 gives us an opportunity to investigate the claims for a correlation between $\gamma$-ray conversion efficiency and pulsar age, over a wider range of characteristic ages. However, it should be cautioned that any conclusions from such a study should take into account the dependency of both efficiency and characteristic age on the spin parameters: i.e.~$\eta = L_\gamma/L_{\rm sd} \propto L_\gamma(P^3/\dot{P})$ and $\tau_{\rm c}\propto P/\dot{P}$. Therefore, other pulsar parameters that depend on the period and/or its time derivatives, for example the surface polar field $B$ or the field strength at the light cylinder radius, may well also exhibit a degree of correlation with the efficiency.

\begin{figure*} 
\begin{center}
\vspace*{10pt} 
\includegraphics[scale=0.45]{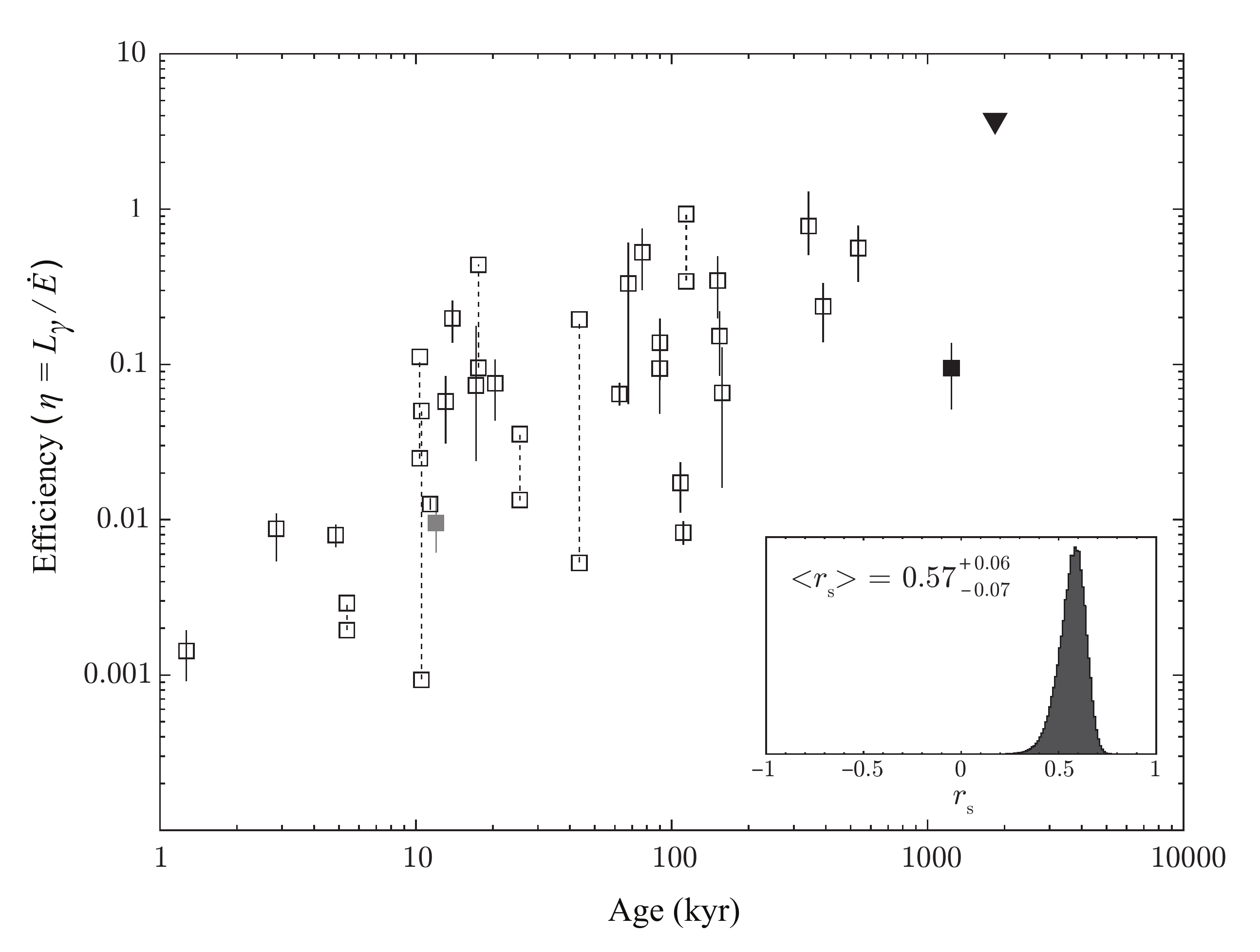}
\caption{\label{fig:efficiency} $\gamma$-ray efficiency plotted against characteristic spin-down age for non-recycled $\gamma$-ray pulsars with available distance estimates (empty squares), and PSR J1836+5925, for which only an upper limit exists (solid triangle). Pulsars with two distance estimates have two markers connected with a dashed line. The solid black square corresponds to the efficiency and age of PSR J2043+2740; the solid gray square corresponds to the same pulsar's efficiency but assuming its age and distance are equal to those of the Cygnus Loop: i.e. $\tau=12$ kyr and $D=540$ pc. All values of $\eta$ in this plot were taken from Abdo et al.~(2010a), except those for PSR J0248+6021 ($\tau_{\rm c}=63$ kyr) and PSR J2240+5832 ($\tau_{\rm c}=151$ kyr), which were taken from \protect\nocite{taa+10} Theureau et al., {\em submitted to ApJ}. The inset plot, at the bottom right corner, shows the distribution of $10^6$ Spearman's rank correlation coefficients, $r_{\rm s}$, from an equal number of Monte-Carlo realisations of the data based on the measured efficiencies and their errors. The median of the distribution is $r_{\rm s}=0.57^{\, +0.06}_{\, -0.07}$, where the uncertainties correspond to the 1$\sigma$ asymmetric errors.}
\end{center}
\end{figure*}

The $\gamma$-ray luminosity of a pulsar can be written as $L_\gamma = 4 \pi f_\Omega G_{\mathrm{> 0.1\ GeV}} D^2$. Based on the measured $\gamma$-ray luminosities of non-recycled pulsars recently detected with {\em Fermi}, Abdo et al.~(2010a)\nocite{aaa+10c} calculated the corresponding $\gamma$-ray efficiencies under the assumption of $f_\Omega=1$ and using distance estimates based either on the pulsar DM or known kinematic properties and supernova-remnant associations. A scatter plot of the $\gamma$-ray efficiency versus the characteristic age of non-recycled $\gamma$-ray pulsars with available distances is shown in Fig.~\ref{fig:efficiency}. In addition, the plot includes an upper limit on $\eta$, denoted with a solid triangle, corresponding to the efficiency of PSR J1836+5925. It should be noted that the efficiency of PSR J1836+5925 was recently estimated by Abdo et al.~(2010d)\nocite{aaa+10b}, assuming $\eta\propto 1/\sqrt{L_{\rm sd}}$. However, we did not use their estimate in this work, because such an assumption introduces an artificial correlation between the pulsar's spin parameters and the $\gamma$-ray efficiency. 

Despite the large error bars on $L_{\gamma}$, Fig.~\ref{fig:efficiency} reveals a correlation between $\eta$ and $\tau_{\rm c}$, with older pulsars having on average higher $\gamma$-ray efficiencies compared to younger ones. The Spearman's rank coefficient for our data set is $r_{\rm s}=0.57^{\, +0.06}_{\, -0.07}$, which implies a positive correlation between efficiency and age. The probability of chance correlation given that value of $r_{\rm s}$ is only $0.11^{\, +0.46}_{\, -0.09}\%$. 
 
In the plot of Fig.~\ref{fig:efficiency}, the data point corresponding to the highest characteristic age --- without being an upper limit --- is that of PSR J2043+2740 and is denoted by a solid black square. It is notable that the efficiency of PSR J2043+2740 is more comparable with that of pulsars that are at least 10 times younger, while the next three youngest pulsars after PSR J2043+2740, being roughly half as young, have efficiencies that are 2--6 times higher. For a number of pulsars in Fig.~\ref{fig:efficiency}, we have two distance estimates: from the pulsar DM and the NE2001 model, and from and independent method; the latter is based either on the Doppler shift of the HI lines of objects associated with the pulsar, combined with a rotation model for the Galaxy, or on trigonometric parallax, or on estimates by other means (\nocite{aaa+10c}Abdo et al.~2010a). For those pulsars, we have assigned two markers connected with a dashed line, each corresponding to the distance estimate from either method. For PSR J2043+2740, we can suppose that its association with the Cygnus Loop is true: this in turn would imply that its distance is $\approx 540$ pc and, moreover, that its true age is roughly 12 kyr instead of 1.2 Myr. Therefore, if the association is true, then the marker for PSR J2043+2740 should be shifted to the position of the solid gray square in Fig.~\ref{fig:efficiency}, which would make the efficiency of this pulsar more consistent with the observed overall trend shown by the rest of the data points.  

There are indeed a handful of examples for which a pulsar--remnant association has led to a significant revision to the pulsar's age: the case of PSR J0538+2817 and S147 has already been mentioned in the introduction; other examples include the 65-ms X-ray pulsar PSR J1811$-$1925, whose association with the Galactic SNR G11.2$-$0.3 suggested an age that was a factor 12 smaller than the spin-down age (\nocite{krv+01}Kaspi et al.~2001), and PSR B1951+32 --- one of the original 6 $\gamma$-ray pulsars detected with EGRET --- with a characteristic age that is more than 1.5-times higher than the age of its birth site, the SNR CTB 80 (\nocite{mgb+02}Migliazzo et al.~2002). The reason for such discrepancies between $\tau_{\rm c}$ and the true age of a non-recycled pulsar is that the calculation of the characteristic spin-down age assumes in all cases that the pulsar was born with an initial period, $P_0$, that is negligible compared to that observed today. Evidently, in the abovementioned cases, such an assumption does not hold, and those pulsars must have been born with a $P_0$ very close to the period observed today. Hence, in the case where $P_0\ll P$ does not hold, the true age of the pulsar should be calculated from $\tau=P/[(n-1)\dot{P}]\times[1-(P_0/P)^{n-1}]$, where $n=2-P\ddot{P}/\dot{P}^2$ is the pulsar braking index. If $P_0\ll P$, $\tau$ can be approximated with $\tau_{\rm c}$. Under the usual assumption of $n=3$, for pure-dipole magnetic braking, one can calculate $\tau$ for a range of $P_0$ values. By simply equating the true age of PSR J2043+2740 with the upper limit on the age of the Cygnus Loop, one finds that the only possibility of association with the remnant is if the pulsar was born with $P_0\approx 95.6$ ms. 

A conclusive remark based on the above is that there is certainly some support for the pulsar--remnant association on the basis of $\gamma$-ray energetics. Nevertheless, it should be recognised that, as of yet, there exists no firm evidence for such an association and that a chance alignment between PSR J2043+2740 and the Cygnus Loop is as likely. Furthermore, it is important to note that a possible rejection of the above association does not provide indisputable support for the alternative distance to PSR J2043+2740, i.e.~the one derived from the NE2001 model: it is quite possible, given the inaccuracies often associated with distances based on DM (e.g.~Deller et al.~2009\nocite{dtb+09}), that this pulsar is at an even greater distance than predicted by its DM. In fact, PSR J2022+2854, which is $\approx 4\fdg5$ away from PSR J2043+2740 in th sky, is at a DM-distance of 2.7 kpc and has a comparable DM and RM to PSR J2043+2740 (${\rm RM}_{\rm J2022+2854}=-75$ rad m$^{-2}$); while other nearby pulsars, e.g. J2113+2754 at 2 kpc, have a much lower RM (${\rm RM}_{\rm J2113+2754}=-37$ rad m$^{-2}$).

\subsection{Pulsar Orientation, Proper Motion and Birthplace}
The Cygnus Loop region has been searched by a number of authors for the central compact object, but so far there has been no conclusive evidence that led to a positive association (\nocite{trkp94}Thorsett et al.~1994; \nocite{rtj+96}Ray et al.~1996; \nocite{mtt+98}Miyata et al.~1998; \nocite{mot+01}Miyata et al.~2001). 

\begin{figure*}
\begin{center}
\includegraphics[scale=0.8]{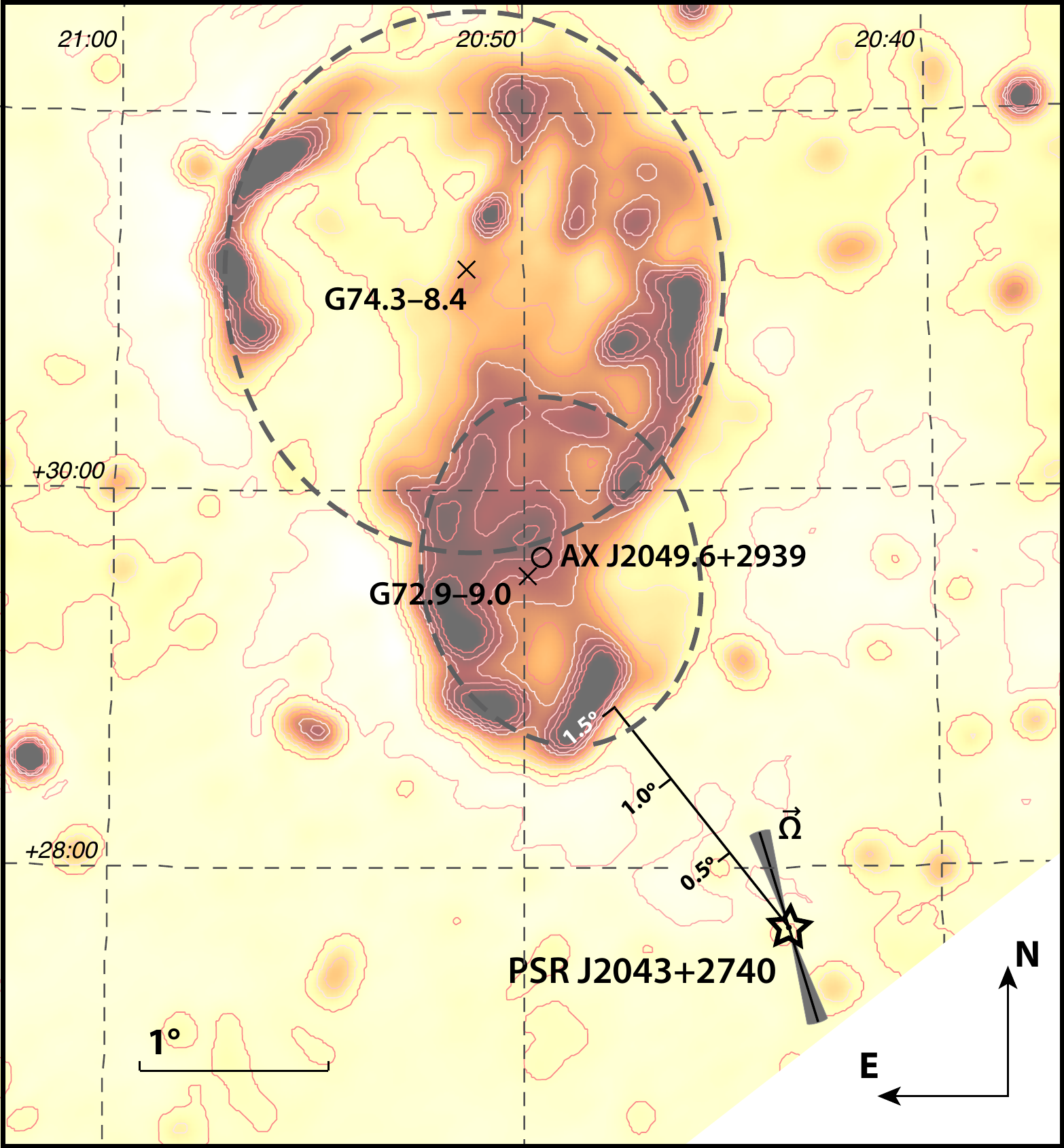}
\caption{\label{fig:cygloop} Radio map of the Cygnus Loop region from data from the Effelsberg 1.4 GHz Medium Galactic Latitude Survey (EMLS; Uyan{\i}ker et al.~1998; Uyan{\i}ker et al.~1999; Reich et al.~2004). In this map, the position of PSR J2043+2740 is shown with a star symbol. At the pulsar's position, we have also drawn the orientation of the pulsar's spin axis (solid, black line), as was derived from radio-polarization data (the shaded, grey area corresponds to the 1$\sigma$ confidence level). In addition, the centers and extents of G74.3$-$8.4 and G72.9$-$9.0, as were calculated by Uyan{\i}ker et al.~(2002), are shown with crosses and gray, dashed lines, respectively. The position of the X-ray compact source, AX J2049.6+2939, discovered by Miyata et al.~(1998) is marked with a circle. Lastly, the map also includes a scale showing the angular separation between the pulsar and the edge of the Cygnus Loop.}
\end{center}
\end{figure*}

Fig.~\ref{fig:cygloop} shows a radio map of the Cygnus Loop remnant relative to the position of PSR J2043+2740; the map was compiled from data from the Effelsberg 1.4 GHz Medium Latitude Survey (EMLS; \nocite{ufr+98}Uyan{\i}ker et al.~1998; \nocite{ufr+99}Uyan{\i}ker et al.~1999; \nocite{rfr+04}Reich et al.~2004). It has been proposed by Uyan{\i}ker et al.~(2002)\nocite{ury+02}, and was later supported with polarization observations at 6 cm (Sun et al.~2006\nocite{srh+06}), that the Cygnus Loop consists of two supernova remnnants: the northernmost G74.3$-$8.4, centered at ($\alpha$,$\delta$)=(20$^{\rm h}$ 51\fm36, +31$^{\circ}$ 3$^{\prime}$), and G72.9$-$9.0, a blowout region in the south-western rim of the Cygnus Loop, centered at ($\alpha$,$\delta$)=(20$^{\rm h}$ 49\fm56, +29$^{\circ}$ 33$^{\prime}$).

Moreover, Miyata et al.~(1998)\nocite{mtt+98} have reported the discovery of a compact X-ray source, AX J2049.6+2939, near the center of G72.9$-$9.0 (see Fig.~\ref{fig:cygloop}). Although it may seem likely that AX J2049.6+2939 could be the central compact source associated with G72.9$-$9.0, follow-up observations with the {\em ASCA} and {\em RXTE} observatories showed significant X-ray variability, which makes the identification of AX J2049.6+2939 as an ordinary, rotation-powered pulsar unlikely (Miyata et al.~2001\nocite{mot+01}).

Given the lack of alternative candidates, the proximity of PSR J2043+2740 ($\approx 1\fdg5$ outside the edge of G72.9$-$9.0) makes this pulsar the only possible prospect for an association. Ultimately, a conclusion to whether PSR J2043+2740 is associated with the Cygnus Loop can be drawn by measuring the pulsar's proper motion. One method of estimating pulsar proper motions is through pulsar timing. We examined 7 years of timing observations with the Lovell telescope, at 1.4 GHz. We fitted 234 TOAs for the pulsar period and its first and second time derivatives. The post-fit timing residuals displayed the typical, long-term (i.e.~``red''), correlated timing signature often seen in the TOAs of non-recycled pulsars: this phenomenon is commonly referred to as ``timing noise'' (see \nocite{hlk+04}Hobbs et al. 2004; \nocite{hlk10}2010). Younger and more energetic pulsars often exhibit a higher amount of timing noise than older ones. PSR J2043+2740 is ranked amongst the pulsars with the highest amount of timing noise, a fact which is demonstrated by this pulsar's high and significant value of $\ddot{\nu}=56.04(57)\times 10^{-24}$ s$^{-3}$ (compare with Table 1 of Hobbs et al.~2010).

The fitting procedure that can be used to derive the pulsar proper motion from timing data only provides correct results when applied to statistically ``white'' data. Therefore, some form of pre-whitening needs to be applied to effectively remove the timing noise mentioned above (e.g.~\nocite{hem06}Hobbs et al.~2006). We attempted the standard technique of fitting harmonically related sinusoids, but given the strength of the timing noise in this pulsar we were unable to pre-whiten the timing residuals sufficiently to obtain a plausible fit for proper motion. 

Another method for inferring the pulsar's proper-motion direction is through our radio-polarization measurements (section~\ref{subsec:poldata}). It has been claimed that there is strong evidence for a correlation between the projected spin-axis orientations of pulsars and their respective velocity-vector directions (Johnston et al.~2005\nocite{jhv+05}; Johnston et al. 2007\nocite{jkk+07}). It has also been suggested that such a correlation --- if there is indeed a physical mechanism that aligns pulsar spin axes with their velocities --- should vanish for old pulsars, as the Galactic gravitational potential will have had enough time to alter the velocities of those pulsars. On the other hand, the young population of pulsars should display a stronger case for alignment. 

The position angle ${\rm PA}_0$ defines the orientation --- relative to our polarization dipoles --- of the plane of linear polarization of the pulsar's emission, at the closest approach of the magnetic pole to our line-of-sight. Assuming that the plane of polarization coincides with the plane defined by the magnetic-field lines and the magnetic axis, ${\rm PA}_0$ should also correspond to the orientation of the spin axis, projected on our field-of-view: i.e.~${\rm PA}_{\rm r}\equiv{\rm PA}_0$. Note that the calculation of the PA from Stokes $Q$ and $U$ can only provide a `headless' vector with $180^\circ$ ambiguity. In addition, as was shown by Backer, Rankin \& Campbell (1975)\nocite{brc75} and Manchester, Taylor \& Huguenin (1975)\nocite{mth75}, pulsar emission can occur in two orthogonal modes: i.e.~the polarization plane can either coincide with the field line--magnetic axis plane or be perpendicular to it. Therefore, there is an additional 90$^\circ$ ambiguity in the projected direction of the spin axis.

If we assume that the scenario of alignment between the spin axis and velocity is true for young pulsars and that PSR J2043+2740 was indeed born within the bounds of the Cygnus Loop --- making it a young pulsar --- then we expect that the spin-axis orientation is, within the measurement errors, directed towards the SNR. Our polarization measurements show that the most favorable case to the above scenario, given the $90^\circ$ ambiguity, gives a  position angle for the spin axis of $17^\circ$ North-through-East. The spin-axis orientation relative to the Cygnus loop is shown in Fig.~\ref{fig:cygloop}, drawn at the pulsar's position. From the figure, it is immediately evident that the spin-axis orientation is significanlty offset from the direction to the centre of G72.9$-$9.0, the offset being $\approx 19^\circ$.
  
A direct implication of the above offset could be that our assumptions are not valid and that PSR J2043+2740 is not associated with the Cygnus Loop. Another possibility is that PSR J2043+2740 was born in the supernova but the velocity vector is not aligned with the pulsar's spin axis: it is true that in the work of Johnston et al.~(2005) the offsets between the velocity vectors and the spin axes show a significant spread. However, from the same work, Johnston et al.~(2005) derive that roughly 90\% of the pulsars have an offset of $<19^\circ$ between the spin-axis orientation and the velocity vector, which makes the measured offset for PSR J2043+2740 an unlikely product of any underlying alignment mechanism leading to the observed distribution of offsets.

In conclusion, the above arguments for the pulsar--remnant association, based on the pulsar orientation and its inferred proper-motion direction, do not appear to favor the Cygnus Loop as the birthplace of PSR J2043+2740. However, all of these arguments are heavily based on uncertain assumptions; a proper-motion measurement is still required for an unequivocal verdict.   

Recently, it was proposed that a large part of the observed timing noise of PSR J2043+2740, as well as that of other pulsars, is due to pulse-shape variations (\nocite{lhk+10}Lyne et al.~2010); those variations appear to be correlated with changes in $\dot{P}$. Moreover, it was suggested that one can fit for such pulse-shape variations and mitigate the large variations in the timing residuals by a significant factor. In particular, PSR J2043+2740 shows major pulse variations that cause 100\% change in the pulse's FWHM on time scales of $\approx 200$ days. Our data set, covering nearly 500 days, is certainly affected by those changes; still, for a proper-motion fit that would use years of data --- as is needed to obtain a precise measurement --- such an effect dominates over the proper-motion signature: i.e.~for PSR J2043+2740, the RMS of the timing residuals over 4,000 days is $>1$ s (e.g.~see Fig.~1 of \nocite{lhk+10}Lyne et al.~2010). In conclusion, such a method is certain to assist proper-motion estimates through pulsar timing for a large number of `noisy' pulsars, such as PSR J2043+2740.

Alternatively, a direct measurement of the pulsar's proper motion can be made with VLBI observations. If the pulsar was born 12 kyr ago in the supernova explosion that created G72.9$-$9.0, then the angular separation between the center of the SNR and the pulsar, i.e.~$\approx {\rm 2\fdg3}$, translates to a transverse velocity of $V_{\perp}\sim 1,770$ km s$^{-1}$, assuming both pulsar and remnant are at 540 pc distance. Indeed, this corresponds to a large value of proper motion ($\approx 690$ mas yr$^{-1}$), with the highest inferred transverse velocities published so far being $V_{\perp}\sim 1,600$ km s$^{-1}$ (\nocite{hllk05}Hobbs et al.~2005). We have performed TEMPO2 simulations that generate fake TOAs having a proper-motion signature corresponding to a pulsar at the position of PSR J2043+2740, with 690 mas y$^{-1}$. Fitting those TOAs with the pulsar's spin parameters but setting the proper motion to zero, reveals the characteristic periodic sine-wave, modulated with time difference from the reference position epoch. The maximum amplitude of the simulated wave over a few thousand days, due to the proper motion, was $\sim 10$ ms (see also Fig.~8.2d in \nocite{lk05}Lorimer \& Kramer 2005). This otherwise large amount of residual RMS is still swamped, for the case of PSR J2043+2740, by the much larger timing noise of this pulsar.  However, if PSR J2043+2740 possesses such a large proper motion, it will be easily measurable with VLBI, which will conclusively confirm or reject the pulsar--remnant association.   

\vspace{0.5cm}
\section{Summary}
The present article presented the results of the analysis of radio and $\gamma$-ray data from PSR J2043+2740, taken with the Jodrell Bank and Effelsberg radio-telescopes, and the {\em Fermi} $\gamma$-ray observatory. Table~\ref{tab:results} gives a summary of the main numerical results from the above analysis.  

We used a radio ephemeris from Jodrell Bank timing observations to fold 14 months of {\em Fermi} $\gamma$-ray data from PSR J2043+2740. The resulting $\gamma$-ray lightcurve clearly shows a double-peaked structure with a significance of $\approx 7\sigma$. Using a bin-independent statistical analysis, we derived a pulse-width of $\approx 0.19$ pulse periods for the leading $\gamma$-ray peak and a significantly narrower width of $\approx 0.06$ pulse periods for the trailing peak. The highest-energy event present in the constructed lightcurve has $E=4.9$ GeV and was found coincident with the bridge emission between the $\gamma$-ray peaks. 

The significant detection of a $\gamma$-ray lightcurve from PSR J2043+2740 allowed us to further characterize the properties of the pulsar's $\gamma$-ray emission: we have measured a phase lag of $\delta\approx 0.2$ pulse periods between the radio and $\gamma$-ray emission; we also measured the phase separation between the major $\gamma$-ray peaks to be $\Delta\approx 0.35$ pulse periods. 

The shape and structure of the $\gamma$-ray lightcurve allowed us to constrain the beam geometry of PSR J2043+2740 using the recently published ATLAS of $\gamma$-ray lightcurves by Watters et al. We only considered the ATLAS results from the two prevailing outer-magnetospheric models, the Two Pole Caustic model and the Outer Gap model. Our assumption that the $\gamma$-ray emission for PSR J2043+2740 comes from the outer magnetosphere was also supported by theoretical estimates based on the highest energy detectable from this pulsar: we estimated that the high-energy emission most likely comes from a distance of at least $1.4R_\star$ from the star's center. 

We combined the geometrical constraints from the $\gamma$-ray data with the results from 21-cm polarization observations of PSR J2043+2740 with the Effelsberg telescope: by fitting the rotating vector model of Radhakrishnan \& Cooke to the absolute polarization angles across the radio pulse, we produced confidence contours constraining the magnetic inclination and viewing angles. The $\gamma$-ray and radio constraints combined place stringent limits on the allowed ranges for these angles. For the Two Pole Caustic model, we determined $\{\alpha,\zeta\}\sim\{52^\circ$--~$57^\circ,61^\circ$--~$68^\circ\}$ to be a valid parameter range. For the Outer Gap model, we found two separate ranges that satisfied the data constraints: $\{\alpha,\zeta\}\sim\{62^\circ$--~$73^\circ,74^\circ$--~$81^\circ\}$ and $\{\alpha,\zeta\}\sim\{72^\circ$--~$83^\circ,60^\circ$--~$75^\circ\}$. 

Given the presence of valid ranges for $\alpha$ and $\zeta$ for both models, we could not exclude either of them based on emission-geometry arguments. However, it was noted that the roughly symmetric shapes of the $\gamma$-ray peaks in the lightcurve of PSR J2043+2740 favor the predictions of the Outer Gap model but are incompatible with the asymmetric peaks predicted by the Two Pole Caustic model.

In addition, we performed a spectral analysis on all $\gamma$-ray events above 0.1 GeV, from the direction of PSR J2043+2740. The best fit to the measured fluxes as a function of photon energy was compatible with a power-law of spectral index $1.1\pm 0.2$ and an exponential cut-off at $0.8\pm 0.1$ GeV. 

Previous claims that pulsars become more efficient at emitting $\gamma$ rays as they age were also investigated in this paper. We plotted the $\gamma$-ray efficiency of PSR J2043+2740 together with those of other non-recycled $\gamma$-ray pulsars as a function of characteristic age. Despite the large scatter present in the data, we saw an evident trend towards higher efficiencies with growing pulsar age. The position of PSR J2043+2740 in our plot is somewhat lower than the expected efficiency based on the average trend of the total sample.

Furthermore, we have also explored the possibility that PSR J2043+2740 was born in the nearby, 12-kyr supernova remnant, the Cygnus Loop. By using the pulsar--remnant association as a postulate, we concluded that the pulsar's $\gamma$-ray efficiency is comparable, within the measurement errors, to that of most other pulsars of similar characteristic age as the age of the Cygnus Loop.

In an attempt to make a conclusive statement about the above association, we used Jodrell Bank timing data from PSR J2043+2740, in order to measure its proper motion through pulsar timing: a proper motion vector pointing away from the center of the Cygnus Loop would unquestionably make the latter the pulsar's bithplace. Unfortunately, due to the high timing noise present in the pulsar's timing residuals, we were unable to fit for proper motion. 

Alternatively, we considered an indirect method of estimating the pulsar's proper-motion direction, based on the claims of Johnston et al.~that young pulsars tend to have their spin axes and proper-motion vectors aligned. Therefore, if PSR J2043+2740 is associated with the Cygnus Loop, it should be young and have a proper motion directed along the spin axis. From the rotating-vector-model fits to the radio-polarization angles, the best determined value of the spin-axis orientation of PSR J2043+2740 was $17^\circ\pm 6^\circ$, measured from North through East. The derived spin-axis direction points $19^\circ$ away from the nearest SNR center, i.e.~that of the Cygnus Loop's southwestern blowout. This result weakens the arguments for an association: it means that either PSR J2043+2740 was not born in the Cygnus Loop or possibly --- although unlikely --- that the velocity vector of this pulsar is not aligned with its spin axis and that the former does in fact point away from the center of the SNR. If the latter is true, our calculations showed that the current angular separation between PSR J2043+2740 and the southwestern part of the Cygnus Loop implies a transverse velocity of $V_{\perp}\sim 1,770$ km s$^{-1}$ for the pulsar. If this pulsar is indeed moving away from the Cygnus Loop at such a high velocity, future VLBI measurements should be able to easily measure it and shed light on this pulsar's connection to the Cygnus Loop.

\vspace{0.5cm}

The \textit{Fermi} LAT Collaboration acknowledges generous ongoing support
from a number of agencies and institutes that have supported both the
development and the operation of the LAT as well as scientific data analysis.
These include the National Aeronautics and Space Administration and the
Department of Energy in the United States, the Commissariat \`a l'Energie Atomique
and the Centre National de la Recherche Scientifique / Institut National de Physique
Nucl\'eaire et de Physique des Particules in France, the Agenzia Spaziale Italiana
and the Istituto Nazionale di Fisica Nucleare in Italy, the Ministry of Education,
Culture, Sports, Science and Technology (MEXT), High Energy Accelerator Research
Organization (KEK) and Japan Aerospace Exploration Agency (JAXA) in Japan, and
the K.~A.~Wallenberg Foundation, the Swedish Research Council and the
Swedish National Space Board in Sweden.

Additional support for science analysis during the operations phase is gratefully
acknowledged from the Istituto Nazionale di Astrofisica in Italy and the Centre National d'\'Etudes Spatiales in France.

The Lovell Telescope is owned and operated by the University of Manchester as part of the Jodrell Bank Centre for Astrophysics with support from the Science and Technology Facilities Council of the United Kingdom.

The authors would also like to give special thanks to Drs.~K.~J.~Lee, W.~Reich and J.~Verbiest for their invaluable contribution and helpful comments, which helped to vastly improve the present article.

\bibliography{journals,modrefs,psrrefs,crossrefs}

\clearpage

\pagebreak

\pagebreak

\pagebreak



\pagebreak

\pagebreak

\pagebreak

\clearpage

\begin{table*}
\caption{Main results from the herein analysis.}
\label{tab:results}
\begin{tabular}{lrll}
\hline
\vspace{-0.35cm} & & & \\
Property & Value (${\rm Error\,}^a$) & Units & Description \\
\vspace{-0.4cm} & & & \\
\hline
\hline
\vspace{-0.35cm} & & & \\
\hspace{0.3cm}{\em $\gamma$-ray profile} & & & \\
\hline
\vspace{-0.3cm} & & & \\
$\phi_1$    				& 0.20(3)  								& periods   								  & Phase of leading $\gamma$-ray peak$^b$  				\\
$\phi_2$    				& 0.55(2)   								& periods   								  & Phase of trailing $\gamma$-ray peak     				\\
$w_1$       				& 0.19(6)         								& periods  			 					  & Width of leading $\gamma$-ray peak		                	\\
$w_2$       				& 0.06(1)         								& periods   								  & Width of trailing $\gamma$-ray peak		                	\\
$\Delta$    				& 0.35(4)      								& periods   								  & $\gamma$-ray peak separation  					\\
$\delta$    				& 0.20(3)      								& periods  								  & $\gamma$-ray--radio peak separation    	\\
\vspace{-0.2cm} & & & \\
\hline
\vspace{-0.35cm} & & & \\
\hspace{0.3cm}{\em $\gamma$-ray energetics} & & & \\
\hline
\vspace{-0.3cm} & & & \\
$\Gamma$    				& 1.1(2)       									& --                           						  & Power-law index      					        \\
$E_{\rm c}$ 				& 0.8(1)      									& GeV   					                          & Exponential cut-off energy 						\\
$F_{\mathrm{> 0.1\ GeV}}$ 		& 2.2(4) 									& 10$^{-8}$ ph cm$^{-2}$ s$^{-1}$           			          & Integral photon flux $>0.1$ GeV       				\\
$G_{\mathrm{> 0.1\ GeV}}$ 		& 1.5(2) 									& 10$^{-11}$ erg cm$^{-2}$ s$^{-1}$                                    	  & Energy flux $>0.1$ GeV        					\\
$L_{\gamma}$ 				& 5.3(24)$^c$ 									& 10$^{33}$ erg s$^{-1}$                                    	  	  & $\gamma$-ray luminosity $>0.1$ GeV       					\\
$\eta_{\gamma}$      			& 0.09(4) 									& 10$^{-11}$ erg cm$^{-2}$ s$^{-1}$                                    	  & $\gamma$-ray efficiency $>0.1$ GeV       					\\
\vspace{-0.2cm} & & & \\
\hline
\vspace{-0.35cm} & & & \\
\hspace{0.3cm}{\em Geometry \& polarization} & & & \\
\hline
\vspace{-0.3cm} & & & \\
$\alpha$ (TPC/OG)    			& 52$^\circ$--~57$^\circ$ / 62$^\circ$--~73$^\circ$ {\em or} \ 72$^\circ$--~83$^\circ$	   &                                                              & Inclination angle ranges$^d$                			\\
$\zeta$  (TPC/OG)    			& 61$^\circ$--~68$^\circ$ / 74$^\circ$--~81$^\circ$ {\em or} \ 60$^\circ$--~75$^\circ$         & 						                  & Viewing angle ranges 						\\
RM          				& $-$92.7(19)    								& rad m$^{-2}$							          & Rotation measure							\\
$\phi_0$        			& 0.032(2) 									& periods	  							  & Minimum approach to the magnetic pole$^e$				\\
${\rm PA}_{\rm r}$        		& 17(6) 									& degrees 								  & Spin axis position angle$^f$					\\
\vspace{-0.3cm} & & & \\
\hline 
\end{tabular}
\\
\\
{\footnotesize $a.$ All errors in parentheses are on the last significant digit of the value to which they refer.} \\
{\footnotesize $b.$ Relative to the radio peak.} \\
{\footnotesize $c.$ The error on $L_{\gamma}$ incorporates the uncertainties in $G_{\mathrm{> 0.1\ GeV}}$ and $D$. All flux calculations assume $f_\Omega=1$.} \\
{\footnotesize $d.$ Each range of $\alpha$ is matched with the corresponding range of $\zeta$, directly beneath it.} \\
{\footnotesize $e.$ Relative to the maximum of the radio emission; positive values denote delay.} \\
{\footnotesize $f.$ Measured North through East.} \\

\end{table*}

\end{document}